\documentstyle[graphicx,12pt]{aipproc}
\advance \textheight by \topskip
\begin{document}

% Macro definitions
\def\nuc#1#2{${}^{#1}$#2}
\newcommand{\Gese}{\mbox{$^{68}$Ge}}
\newcommand{\Geso}{\mbox{$^{71}$Ge}}
\newcommand{\Gaso}{\mbox{$^{71}$Ga}}
\newcommand{\Gesn}{\mbox{$^{69}$Ge}}
\newcommand{\Feff}{\mbox{$^{55}$Fe}}
\newcommand{\Arts}{\mbox{$^{37}$Ar}}
\newcommand{\Be}{\mbox{$^8$B}}
\newcommand{\Bes}{\mbox{$^7$Be}}
\newcommand{\nBe}{\mbox{$\nu_{Be}$}}
\newcommand{\nB}{\mbox{$\nu_{B}$}}
\newcommand{\Cdoon}{\mbox{$^{109}$Cd}}
\newcommand{\Tn}{\mbox{$T_N$}}
\newcommand{\Rnttt}{\mbox{$^{222}$Rn}}
\newcommand{\germ}{\mbox{GeH$_{4}$}}
\newcommand{\nwt}{\mbox{Nw$^{2}$}}

\title{Possibility for Teleportation of Nuclei}

\author
{$^{a}$\,B.\,F. Kostenko\,\thanks{e-mail: kostenko@cv.jinr.ru}
\and $^{a}$\,V.\,D. Kuznetsov
\and $^{b}$\,M.\,B. Miller\,%\thanks{e-mail: iftp@dubna.ru} % $^b$
\and \mbox{$^{b}$\,A.\,V. Sermyagin}
\and and $^{a}$\,D.\,V. Kamanin}
\address{$^{a}$ Joint Institute for Nuclear Research\\%
Dubna Moscow Region 141980, Russia\\
$^{a}$Institute of Physical and Technology Problems\\ P.O.\,Box 39 Dubna Moscow
Region 141980, Russia\\}
\maketitle
\begin{abstract}
Since its discovery in 1993, quantum teleportation (QT)
is a subject for intense
theoretical and experimental efforts. Experimental
realizations of QT have so far been
limited to teleportation of light. The present
letter gives a new experimental scheme
for QT of heavy matter. We show that the standard
experimental technique  used
in nuclear physics may be successfully applied
to teleportation of spin states of atomic nuclei.
  It was shown that there are no theoretical prohibitions
  upon a possibility of a complete
   Bell measurement, so that implementation of all four
   quantum communication channels is at
    least theoretically available. A general
    expression for scattering amplitude of two
    $\frac12$-spin particles was given in the Bell
    operator  basis, and peculiarities of Bell states
    registration are briefly discussed.
\smallskip
\end{abstract}

\section*{Introduction}

Not long ago only science fiction authors ventured
to use a term "teleportation".
However in the last few years the situation
drastically changed. In a landmark work
\cite{1}  a procedure for
teleporting an unknown quantum
state from one location to another
was described. Recent experiments have proved
that this process
can actually happen \cite{2,3}. Now invention of QT
is expected to have a great influence
on the future computation and communication
hardware comparable with
the impact of radio network
on modern technique. It may have
important applications in superfast quantum
computers (theoretical at present) \cite{4}-\cite{7}
as well as in utilizing quantum phenomena to ensure
a secure data transmission
(by means of so-called quantum cryptography) \cite{8}-\cite{10}.
Practical realization of quantum information
processing requires special
quantum gates which cannot be performed through unitary
operations,
but may be constructed with the use of quantum
teleportation for a basis element \cite{11}.
Recently a one-to-one correspondence between
quantum teleportation and dense
coding schemes were established as well \cite{12}.
Besides a relevancy to such applications
as quantum computing, QT is also a new fundamental concept
in quantum physics. Experimental
demonstrations show that QT is an
experimentally achievable technique to
study the phenomenon of quantum entanglement. Indeed, the very
phenomenon of QT appeared to be
possible only due to the
Einstein-Podolsky-Rosen correlations
(see below), which till now are confirmed exactly only
for photons. The same is
true for QT, because  only entangled optical beams
have been so far used to teleport  quantum states of massless matter.

Since quantum information processing involves
material particles such as atoms and ions,
teleportation of heavy matter is considered now
as the next necessary step for obtaining  a complete set of quantum processing
tools \cite{13}-\cite{16}.

We propose here  a new experimental scheme for QT
of heavy matter  based on a standard experimental nuclear physics
technique and expected to be fulfilled
in the nearest one or two years. To the best of our knowledge
other methods require at least ten years to be successful.

\section*{Action-at-a-distance (Teleporting information)}
In 1935 Albert Einstein and his colleagues
Boris Podolsky and Nathan Rosen (EPR) developed a gedanken
experiment to show as they believed a defect
in quantum mechanics (QM)\,\cite{EPR,Bohr}.
This experiment has got the
name of EPR-paradox. An essence of EPR-paradox
is as follows.
There are two particles that have interacted with
each other for some time and have constituted a
single
system. In the QM that system
is described by a certain wave function.
When the
 interaction  is terminated and the particles fled
 far away from each other
 they are as yet described by
 the same wave function.  However individual
 states
 of each separated
  particle are completely unknown. Moreover,
  definite individual properties
  do not exist in principle
  as the QM postulates prescribe.
   It is only after {\bf one} of the particles is
   registered by a detection instrument
   that the states
   arise to existence
   for {\bf both} of them. Furthermore, these
   states are generated  simultaneously
   regardless of    the distance
between the particles at the moment. It looks like
one particle informs instantly the other  of its state.

The real (not just "gedanken") experiments on
teleportation of information of this type,
or "a spooky-action-at-a-distance", as A. Einstein
called it,
were carried out only 30-35 years later, in the
seventies-eighties\,\cite{Aspect,Clauser}. Experimenters,
however, managed to achieve full and
definite success only for photons,
though attempts to perform experiments with
atoms\,\cite{atoms}
and protons were also undertaken\,\cite{proton}.
For the case of two photons the experiments were
carried out for various distances between
them at the moment of registration,
and the EPR-correlations  were shown to survive
up to as large distances as  more than
ten kilometers\,\cite{10 kilometers}.
In the case of protons, an experiment
was carried out only for much less distances of about a
few centimeters and the condition of causal
separation, $\Delta x>c\Delta t$,
was not met. Thus it was not fully persuasive,
as have been recognized by the
authors of the work\,\cite{proton} themselves.

\section*{Teleporting photon-quantum state (or the light
quantum itself?)}

The next step in this direction that suggested itself was
not merely the "action-at-a-distance",
but  transmission   of  a quantum state
from one  quantum object to another .
Namely, this process was called QT.
In spite of the successful
EPR-effect experiments it was
until   recently   even this kind
of teleportation was believed to be impossible at all.
At first sight it  seems as Heisenberg
uncertainty principle would forbid the
very first  step of the  teleportation
which was meant as an extraction of complete information
about the inner properties of a quantum
object to be teleported.
But it cannot be done because of an impossibility
to measure simultaneously  exact
values for the so-called complementary
variables of a quantum microscopic object
(e.\,g. spatial coordinates and momenta).
Nevertheless, in 1993, a group of physicists
(C.\,Bennet and his colleagues) managed
to get round this
difficulty\,\cite{1}.
 They showed that measurement of full quantum information
 is not necessary for
 quantum states transferring from one
 object to another. Instead,
 it was proposed to create that
 a  so-called EPR-channel of communication
  on the basis of EPR-pair of two
 quantum particles.
Let it be photons B and C, shown in Fig.\,1.
 After they have interacted in a way to form a
 single system,  decaying afterward,
 the photon B is directed to the "point of departure",
 where it meets A  within a  registration
 system. The system is arranged in a mode
 (see below) to "catch" only  those events
 which leave no choice to C but
 to take a state that A had
 initially (before its
 interaction with B in the detector at
 the "point of departure").
 This experimental technique is very fine but well known to
 those skilled in the EPR-art.
%The conservation
% laws of general physics are the basis of the
% procedure realizing the system with a given selective
% sensitivity.
% The end of all these manipulation is that particle C gets
% something from A. It is only the quantum state.

What is important from the principal point of view,
it is  "disappearing" of A in the place,
notified in  Fig.\,1 as "Zone
of scanning" (ZS). Indeed,
interaction of B and A
destroys the A photon, in a sense
that none of the two photons outgoing from
ZS  has definite properties of A.
They constitute  a new
pair of photons, which only as a whole
has some quantum
state, and the individual components of the pair
are deprived of this
property. Therefore, in some sense the photon A
really disappears at ZS. Exactly at the
same moment the photon C obtains the properties A had in the
beginning. Once it is happened, in view of the
principle of identity
of elementary particles, we can say
that A, disappearing at ZS, reappears
at another location. Thus, the quantum
teleportation is accomplished.

\begin{figure}[ht]
\includegraphics[width=\textwidth]{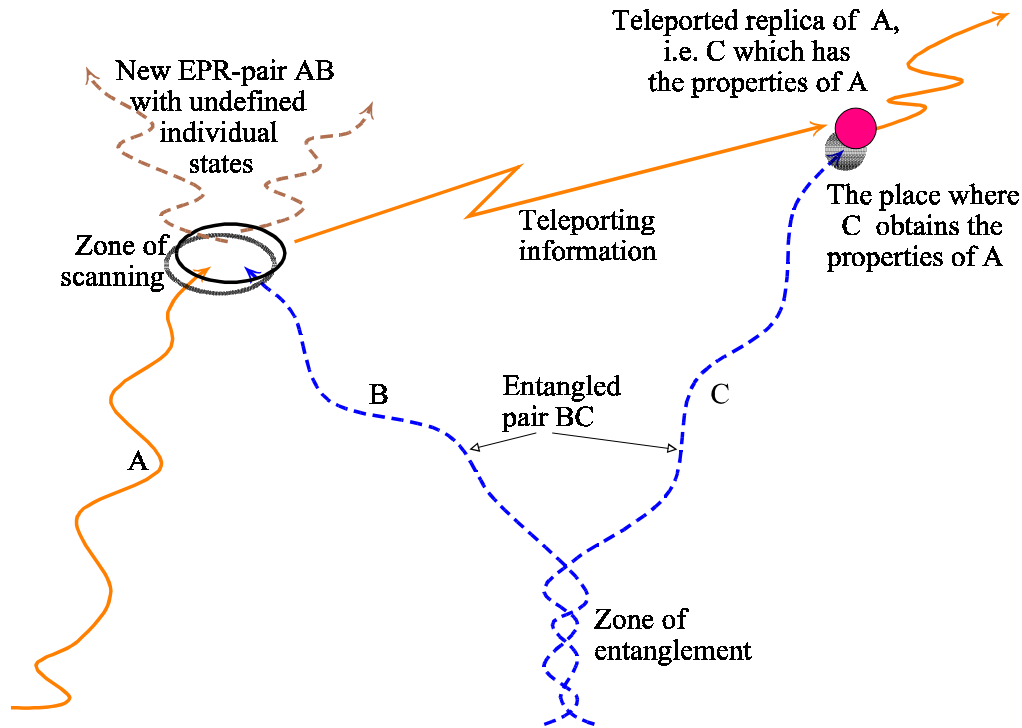}
\caption{Illustration of a general idea how the
teleportation can be realized. Here A is
a  photon we want to pass to a destination place,
B and C, representing an EPR-pair
 of photons, constitute
a so-called quantum transmission channel.
As a result,
definite properties of A are destroyed completely
at the zone of scanning, and at another place
we have the photon C with  the properties A  had
just before it met intermediary object B ("vehicle").
Note that the vehicle first contacts
the C photon to which
the "cargo" has to be
transported, and only later
it calls A to take the cargo from it!}
\label{cntrate}
\end{figure}

 This process has several paradoxical features.
 In spite of the absence of
  contacts between objects (particles, photons) A and C,
  A manages to pass its properties to C.
  It may be arranged in such a way  that
  the distance from A to C  is large enough
  to prevent any causal  signals between them!
  Furthermore,
  in contrast to the transportation of ordinary
  material cargo,   when a delivery vehicle   first
  visits   the   sender to collect a cargo from it,
  quantum properties
  are delivered in a backward fashion. Here the photon
  B plays a role of the delivery vehicle, and one can see
  that   B first interacts with the recipient (C photon)
  and only after that it travels   to the sender (A)
  for the "cargo".

Finally, to  reconstruct
initial object completely it is necessary
%to fix a time
%moment when the interaction of A and B occurred
%(the moment of the arrival of the "vehicle"
to inform a receiver  at the
destination about a result of the measurement in ZS.
%to the departure "station" after it visits the
%recipient),
This allows him to accomplish  processing of
quantum signal (incoming with the particle C) in a due manner.
%The task of recording the moment
%of (A-B)-interaction and
%using it in the data
%analysis together with
%the information transmitted
%by a quantum EPR-channel requires
Therefore,  one more channel of communication is needed
for an ordinary or classical information
transmission. Only receiving a message
(using the classical communication line) that
A and B  form a {\bf new EPR--pair with zero total
spin}, an observer at destination may be sure
that the properties of C
are identical to those of A before  teleportation.
In the case when A~$+$~B system has non a zero total spin,
some additional transformation of quantum
signal is needed (see below).

\bigskip
The new idea was immediately recognized as an
important one and several  groups of experimenters set
to implement  it concurrently.
Nevertheless, it took more than four years to overcome
all technological obstacles
on the way\,\cite{2,3}.
That was because such experiments,
being the records, are always a  step  beyond
the limits of experimental state of the art
achieved before.

\section*{Start with protons}

%An analysis of the problem carried out
%by authors of the present
%experimental project
% which is now in a stage of preparation
%takes them to a conclusion
The purpose of this paper is to show
that  experimental setups and instruments
developed for conventional  nuclear-physics studies
%(high-current  accelerators
%of protons and heavier nuclei,
%liquid\cite{Liquid} and
%polarized\cite{Polarized} hydrogen
%targets,
% multi-parameter near $2\pi$-geometry --
% i.\,e. semi-spherical
% aperture -- facility for
%particle detection by a name of
%"Fobos" at Nuclear Reaction Laboratory of the Joint Institute for
%Nuclear Research\cite{FOBOS}),
allow one to design
a new way of performing
non-zero  mass matter teleportation,
with a   prospect to implement  the project
in a rather short time. For example,
in accordance with our estimates,
teleportation of  protons
could be achieved in one or two years.
%and it would
%take some more time to prepare the
%teleportaton of more heavy nuclei, e.g. $^3$He.
%............  ................  ..............
%.............  ...............  ...............

\begin{figure}[ht]
\includegraphics[width=\textwidth]{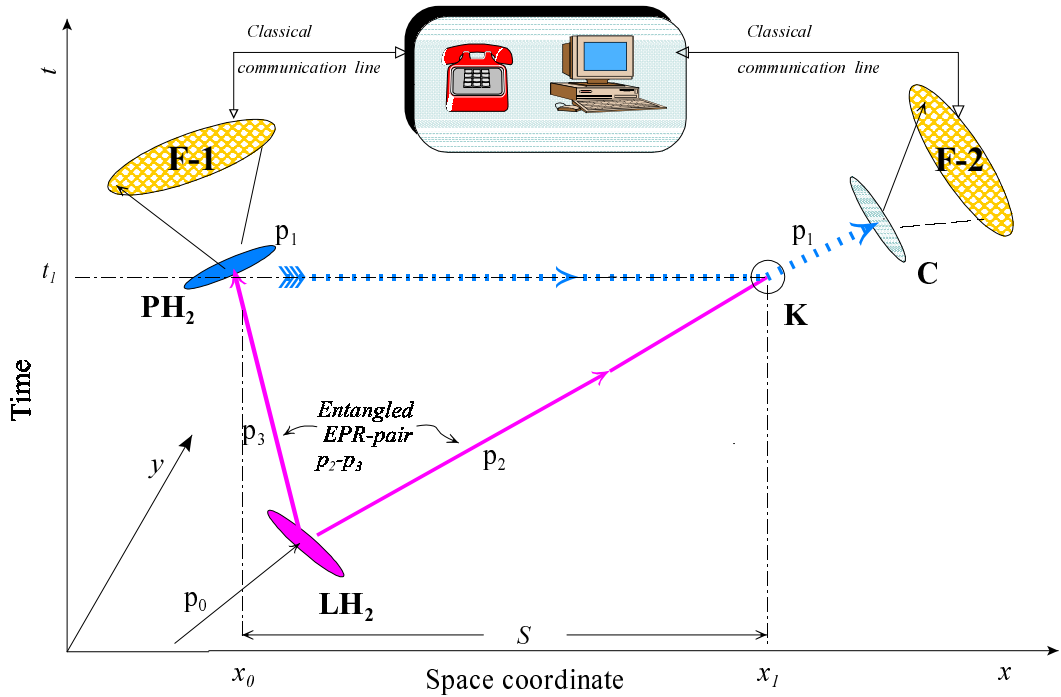}
\caption{Layout of  experiments on proton teleportation.
Here p$_0$ is initial proton from
 the accelerator,
LH$_2$ is a liquid hydrogen target, which
may be also replaced by ordinary polyethylene
(CH$_2$) foils, protons
p${_2}$ and p${_3}$ constitute an entangled EPR-pair,
PH$_2$ denotes
polarized  hydrogen target, C is a carbon target which
operates  as an analyzer of the proton polarization
using the left-right asymmetry of scattering,
F-1 and F-2 are large-aperture position-sensitive
particle detectors (the so-called Fobos-facilities). Proton
spin-state is being teleported from the PH$_2$ target
placed at x$_0$ to the point x$_1$.
%It can be arranged so  that no   signal from x$_0$
%has enough time to reach point x$_1$ before p$_2$
%obtains properties    of p$_1$ at a
%moment t$_1$.
%That fact is justified by the detection system F-1/F-2
%connected with a data-processing center by
%usual communication lines.
K is a point where the spin of p${_2}$ gets a definite
orientation  (which is just the same
that  one of the protons p$_1$ in the  PH$_2$ target had
before the scattering of p${_3}$ from it).
The proton p$_1$ losses its definite quantum
state, as it forms   a new
EPR-pair together with the scattered proton  p${_3}$.
The role of classical communication channel including
a data-processing center is explained in the text.}
\label{prod}
\end{figure}

In  Fig.\,2,  the layout of an experiment
on teleportation of spin states of protons from a
polarized  PH$_2$ target into the point of destination
(target C) is shown. A proton beam p$_0$ of a suitable
energy within the range 20-50
MeV bombards the  LH$_2$ hydrogen target
\cite{Liquid}. According to
the known experimental data, the scattering in
the  LH$_2$ target onto the direction of the second target
(corresponding the angle
$\theta\simeq 90^\circ$ at the c.m.)  occurs
within an acceptable accuracy through the
singlet intermediary state
\,\cite{proton}. Thus, the outgoing protons p$_2$ and p$_3$
form the two-proton entangled system  fully analogous
to the EPR-correlated photons used
%for transmitting information via the quantum
%communication channel
in the experiments
on the teleportation of massless matter,
as it was discussed in the preceding section. At this
moment the system is in a state
$$
|\Psi_{23}\rangle=\frac{1}{\sqrt{2}}\left(|\uparrow_2\rangle|\downarrow_3\rangle-
|\downarrow_2\rangle|\uparrow_3\rangle\right).
$$

One of the scattered protons,
p$_2$, then travels to the point of destination
(the target-analyzer C), while the other,
p$_3$, arrives to a point where  teleportation
is  started, i.\,e., to
PH$_2$-target. The last one is used as  a source
of particles  to be teleported.
Therefore, protons within this target play
the same role asthe  photons A in the  above section.
But there are two features differentiating
the case of protons from the photon one.
First, the protons p$_1$ are within the motionless target
(and thus they are motionless themselves)
with a much more proton density; besides, the protons within the
 PH$_2$ target
have quite definite quantum state, determined
by a direction of polarization,
$$
|\phi_1\rangle=a|\uparrow_1\rangle+b|\downarrow_1\rangle
$$
which could be oriented accidentally and, thus,
unknown to the experimenters.

In the case,
when the scattering in the
polarized  PH$_2$ target occurs in the same
kinematics conditions as in the
LH$_2$ target (i.\,e., at the c.m. angle
$\theta\simeq 90^\circ$),
the total spin of the particles p$_1$
and p$_3$ also must be equal to zero after collision.
To detect the events, a removable circular module F-1
of the facility "Fobos" is supposed to be used
\cite{FOBOS}. Due to this fact,
the detection efficiency is hoped to be much enhanced.
If all the above conditions are provided,
the protons reaching a point K will suddenly receive
the same spin projections as the protons in
the polarized  \cite{Polarized}
 PH$_2$ target.
Indeed, using a so-called Bell's basis,
$$
|\Psi^{(\pm)}_{13}\rangle=\frac{1}{\sqrt{2}}\left(|\uparrow_1\rangle|\downarrow_3\rangle
\pm|\downarrow_1\rangle|\uparrow_3\right),
$$
$$
|\Phi^{(\pm)}_{13}\rangle=\frac{1}{\sqrt{2}}
\left(|\uparrow_1\rangle|\uparrow_3\rangle
\pm|\downarrow_1\rangle|\downarrow_3\right),
$$
the state of three-particle system before the last scattering may
be written in the form
$$
  \begin{array}{l}
|\Psi_{123} \rangle = |\phi_1 \rangle |\psi_{23} \rangle =
\displaystyle{\frac12}
[\; |\Psi^{(-)}_{13} \rangle
(\;\;a |\uparrow_2 \rangle + b |\downarrow_2 \rangle ) +
|\Psi^{(+)}_{13} \rangle
(\;\;a |\uparrow_2 \rangle \;- b |\downarrow_2 \rangle ) \;+
\\
\qquad \qquad \qquad \qquad \; \; \;
+\;\;|\Phi^{(-)}_{13} \rangle
(-a |\downarrow_2 \rangle - b |\uparrow_2 \rangle ) +
|\Phi^{(+)}_{13} \rangle
(-a |\downarrow_2 \rangle + b |\uparrow_2 \rangle )\;] .
  \end{array}
$$
The last scattering and measurement with F-1 select from this
state the term containing $|\Psi^{(-)}_{13}\rangle$,
and therefore the state of the particle 2 will be
$a|\uparrow_2\rangle+b|\downarrow_2\rangle)$.
%It is easily seen that  teleportation of the spin
%states from  PH$_2$-target
%to the recipient p$_2$ really takes place at the point K.
Thus, if the coincidence mode of the detection is provided
via any classical channel, then a  strong
correlation has to take place between polarization
direction in the PH$_2$ target and the direction of the
deflection of  p$_2$-protons scattered in the carbon target C.
Here the carbon foil C plays a  role of the polarization analyzer
, i.\,e.,
one measures the asymmetry of the left-right counting rates
to determine a spin state orientation of p$_2$ before
the scattering \cite{analyzer}.

%The second module  of "Fobos", F-2 at the picture,
%crowns the procedure of teleportation,
%as it indicates the proton
%scattering direction in the
%carbon target C, hence
%its polarization.

In particular, if one succeeds  to make a distance between
the detectors F-1 and F-2 to be sufficiently large
and the difference between the moments of registration
in  F-1 and F-2 to be short enough,
then it will be
possible to meet the important criteria of
the causal independence between the events of the
"departure" of the quantum state from
PH$_2$ target and "arrival" of this "cargo" to the recipient
(proton p$_2$) at the point K.
The measurements consist of
recording signals
entering two independent
but strictly synchronized  memory devices
with the aim to select
afterward those events alone that for sure appeared to be
causal separated.
Thus,  experimental setup shown in Fig.~2 also allows one,
at least in principle,
to fill the gap in verification of the EPR-effect for
heavy matter.
%, for even the most swift signal
%(the light) could not connect them.

%To prevent any exchange
%of signals
%between the points PH$_2$ and K it needs to choose
%appropriately
%proportions of some time and space segments, indicated
%in Fig.2.
%Namely, we have to obtain $S>ct_{12}$,
%where $t_{12} =|t_{F1}-t_{F2}|$.
%Here $t_{F1}$ and $t_{F2}$ are
%moments of registration of signals
%from the corresponding detectors F-1 and F-2
%(their arrival at the
%data collection-processing center).
%For simplicity we neglected
%a time of flight of the protons from K to
%C, and from PH$_2$ and C-targets to the detectors F-1 and F-2,
%respectively.

\section*{General Consideration}
In the experiments that were carried out until now it was
managed to use only one quantum information
transmission channel  corresponding to registration
of Bell's state $|\Psi^{(-)}_{13}\rangle$.
Is it possible to involve  other channels
utilizing the states
$|\Psi^{(+)}_{13}\rangle$, $|\Phi^{(-)}_{13}\rangle$
and $|\Phi^{(+)}_{13}\rangle$?
To answer this question let us consider a general
expression for scattering amplitude of two particles,
not necessarily  identical ones,
with the spin value $\frac12$\,
\cite{Landau},
$$
\begin{array}{c}
\hat{f}=A+B(\vec{S_1}\cdot\vec{\lambda})(\vec{S_2}\cdot\vec{\lambda})
+C(\vec{S_1}\cdot\vec{\mu})(\vec{S_2}\cdot\vec{\mu})
+D(\vec{S_1}\cdot\vec{\nu})(\vec{S_2}\cdot\vec{\nu})\\
+E((\vec{S_1}+\vec{S_2})\cdot\vec{\nu})
+F((\vec{S_1}-\vec{S_2})\cdot\vec{\nu}).
\end{array}
$$
Using  a relation
$$
(\vec{S_1}\cdot\vec{n})(\vec{S_2}\cdot\vec{n})=
\frac{1}{2}\left[((\vec{S_1}+\vec{S_2})\cdot\vec{n})^2-
\frac12\right] ,
$$
 in the case of the coordinate system to be fixed
 for a definiteness  in the following way
$$
\vec{\lambda}\parallel\vec{x},\mbox{\quad}\vec{\mu}\parallel\vec{y},\mbox{\quad}
\vec{\nu}\parallel\vec{z},
$$
the expression for $\hat{f}$ can be represented in the form
$$
\hat{f}=A+\frac{B}{2}\left[S^2_x-\frac12\right]
+\frac{C}{2}\left[S^2_y-\frac12\right]
+\frac{D}{2}\left[S^2_z-\frac12\right]+ES_z-Fs_z,
$$
where
$$
\vec{S}=\vec{S_1}+\vec{S_2}\,,\mbox{\qquad  }
\vec{s}=\vec{S_1}-\vec{S_2} .
$$
The scattering operator $\hat{f}$ can  be now expressed
in terms of the Bell's state transition  operators
making use of the following formulas
$$
\begin{array}{l}
\smallskip
S_x= \;\;\;|\Psi^{(+)}\rangle\langle\Phi^{(+)}|+
|\Phi^{(+)}\rangle\langle\Psi^{(+)}|,\\
\smallskip
S_y=i\left[|\Psi^{(+)}\rangle\langle\Phi^{(-)}|-
|\Phi^{(-)}\rangle\langle\Psi^{(+)}|\right],\\
\smallskip
S_z= \;\;\;|\Phi^{(-)}\rangle\langle\Phi^{(+)}|+
|\Phi^{(+)}\rangle\langle\Phi^{(-)}|,\\
\smallskip
%s_x=-\left[|\Phi^{(-)}\rangle\langle\Psi^{(-)}|+
%|\Psi^{(-)}\rangle\langle\Phi^{(-)}|\right],\\
%\smallskip
%s_y=i\left[|\Phi^{(+)}\rangle\langle\Psi^{(-)}|-
%|\Psi^{(-)}\rangle\langle\Phi^{(+)}|\right],\\
s_z= \;\;\;|\Psi^{(+)}\rangle\langle\Psi^{(-)}|+
|\Psi^{(-)}\rangle\langle\Psi^{(+)}|
\end{array}
$$
and   a decomposition of the unity $
\hat{\bf{1}}=\hat{P}_{\Psi-}+\hat{P}_{\Psi+}+
\hat{P}_{\Phi-}+\hat{P}_{\Phi+}$.
As a result one obtains
\begin{equation}\label{f}
  \hat f = a \hat P_{\Psi-} + b \hat P_{\Psi+}
  +c \hat P_{\Phi-} + d \hat P_{\Phi+} +
  E S_z +F s_z,
\end{equation}
where
$$
a = A - \frac{B+C+D}{4}, \qquad b = a + \frac{B+C}{2},
%$$
%$$
\qquad
c= a+ \frac{C+D}{2}, \qquad d = a + \frac{B+D}{2} .
$$
In the case $E=F=0,$ expression (1) is a usual spectral
decomposition for the operator $\hat f$, which can be
interpreted then  as a quantum observable corresponding to
measurement of one of the Bell's state.  Therefore,
to register a definite Bell's state one has to find
such experimental conditions at which all coefficients but
one of a,b,c, or d in the expression (\ref{f}) turn into
zero. For these purposes, the type and energy of colliding
particles, as well as the  angle  which scattered particles
are recorded at, could be altered. Since the number of
necessary conditions formulated above is less than the
number of free coefficients in (\ref{f}), it is clear that
registration of each Bell's state is possible at least
theoretically.

Directions which spin projections of the scattered
particles should be measured along for
detecting the states $|\Psi^{(+)}\rangle$, $|\Phi^{(-)}\rangle$ and $|\Phi^{(+)}\rangle$
form three orthogonal spatial vectors. It follows from
the  relations
$$
|\Psi^{(+)}\rangle=\vec{e}_1\,,\mbox{\quad
}|\Phi^{(\pm)}\rangle=
\frac{1}{\sqrt{2}}(\vec{e}_2\pm\vec{e}_3),
$$
where $\vec{e}_i$ are orthonormalized
states with the definite values of the spin and
its projections,
$$
\vec{e}_1=|1,0\rangle\,,\mbox{\quad}\vec{e}_2=
|1,1\rangle\,,\mbox{\quad}\vec{e}_3=|1,-1\rangle\,,
$$
which transform in accordance with  3-vector
representation of the rotational group.
It is clear that  spatial rotations at the angle
$\frac{\pi}2$, corresponding to
$\vec{e}_i\rightarrow\pm\vec{e}_j$,
represent the group of permutation for the Bell's states
considered (putting aside an unimportant phase factor -1).
Thus the possibility of registration of $|\Psi^{(+)}\rangle$
state also opens the way  to register  two other states
$|\Phi^{(+)}\rangle$, $|\Phi^{(-)}\rangle$  by means of
change on $\frac{\pi}{2}$ of the direction along which the
spin projection is measured .

For identical spin 1/2 particles the scattering operator
(\ref{f}) has some additional symmetries, so that in c.m.s.
one has
$$
a(\theta) = \; a(\pi - \theta), \qquad
b(\theta) = - b(\pi -\theta),
$$
$$
c(\theta)= -c(\pi - \theta), \qquad
d(\theta)= -d(\pi -\theta),
$$
$$
E(\theta)= \; E(\pi-\theta), \qquad
F(\theta)= F(\pi - \theta).
$$
For nucleon-nucleon scattering we have $F\equiv 0$ as
total spin squared of such a system is conserved and the
last two terms in (\ref{f}) describe transitions
between Bell's state with different $\vec S^2$.
Thus, e.g., for two identical nucleons at
$\theta=\frac\pi2$ one obtains
$$
\hat{f}=a\hat{P}_{\Psi^-}+E\left[|\Phi^{(-)}\rangle\langle\Phi^{(+)}|+
|\Phi^{(+)}\rangle\langle\Phi^{(-)}|\right] .
$$

Experimental identification of Bell's states
$|\Psi^{(-)}\rangle$ and $|\Psi^{(+)}\rangle$ is
rather simple  due to the characterization of these states
by the definite values
of total spin and its projections $(|\vec{S}|=0$, $S_z=0$,
and $|\vec{S}|=1$, $S_z=0$, respectively).
The result of  spin projection
measurement for the particles 1 and 3
%in the state  $|\Psi^{(-)}\rangle$
is
$$
S_{z1}=\pm \; \frac12\,,\mbox{\quad  }S_{z3}=\mp \; \frac12
$$
for any choice of z axis direction, provided their
initial state is $|\Psi^{(-)}\rangle$.

For particles in the $|\Psi^{(+)}\rangle$ state such
correlations take place only if the spin projections
are measured along a definite axis $\vec n$.
If the axis of measuring is deflected
at an angle $\theta$ from this direction the
probability to have $S_{z1} + S_{z3} =0$ will
decrease  as $\cos^2\theta$. One may expect that
at the energies considered,
there is a scattering angle interval corresponding to
$l=1$ and, therefore, to the $|\Psi^{(+)}\rangle$
final state of two protons.

It  seems  more difficult to identify  states
$|\Phi^{(-)}\rangle$ and $|\Phi^{(+)}\rangle$.
In this case, it is necessary first to find out a direction
$\vec n^{\; \prime} $ (which is perpendicular to $\vec n$)
for which measurements of
spin projections give either
$S_{z1} =\frac12$ and $S_{z3}=\frac12$ or
$S_{z1}=-\frac12$ and $S_{z3}=-\frac12$
with the same probability $p=0.5$.
Now measurement of the spin projection of the particle 2
allows one to determine what of two two possible
states,  $|\Phi^{(-)}_{13}\rangle$
or $|\Phi^{(+)}_{13}\rangle$,
the scattering has really occurred into.

\section*{Conclusion}
Referring to the principle of identity of elementary
particles of the same sort with the same
quantum characteristics, i.\,e., the protons in our case,
we can say that protons from a
polarized target PH$_2$ are
transmitted to the destination point C (through the point K). Thus,
in the nearest future, teleportation of
protons can  come from the domain of dreams and fiction
 to the reality in physicists' laboratories.

\bigskip
We wish to thank I.Antoniou, F.A.Gareev, V.V.Ivanov,
O.A.Khrustalev, G.P.Pron'ko,

and V.V.Uzhinsky for helpful
discussions and support.

\bigskip

The work was supported in part by  the
Russian Foundation for Basic Research, project

Nr. 99-01-01101.

\end{document}